# Scholarly Communications of Bharathiar University on Web of Science in Global Perspective: A Scientometric Assessment

**Muneer Ahmad[1*], Dr. M Sadik Batcha[2]**

[1]*Research Scholar, Department of Library and Information Science, Annamalai University, Annamalai Nagar, Tamil Nadu, India*
[2]*Professor and University Librarian, Annamalai University, Annamalai Nagar, Tamil Nadu, India*

***Corresponding Authors:** Muneer Ahmad, Research Scholar, Department of Library and Information Science, Annamalai University, Annamalai Nagar, Tamil Nadu, India, Email: muneerbangroo@gmail.com*

## ABSTRACT

*Bharathiar University is one of the esteemed university and most vibrant in Tamil Nadu. The University was awarded "A" grade by the National Assessment and Accreditation Council (NAAC) of the University Grants Commission and it has been ranked amongst top 50 Universities in India by the survey taken by the prominent English magazines India Today and The Week in 2013 respectively. The present study has been undertaken to find out the impact of research produced and publication trends of the university during 2009 to 2018. The study merely focuses on year-wise research output, citation impact at local and global level, prominent authors and their total output, top journals of publications, collaborating countries, and most contributing departments of Bharathiar University. The 10 years' publication data of the university indicate that a total of 3440 papers have been published from 2009 to 2018 receiving 38104 citations with h-index as 68. In addition to this scientographical mapping of data is presented through graphs using VOS viewer software mapping technique.*

## INTRODUCTION

Named after the great national poet, Subramania Bharathiar, Bharathiar University is a state university in Coimbatore, Tamil Nadu with a motto "Educate to Elevate". The university was incepted in 1982 under the provision of Bharathiar University Act, 1981 (Act 1 of 1982) and was accepted by the University Grants commission in 1985.The university is in the foothills of Marudamalai road, Coimbatore and has jurisdiction over the districts of Coimbatore, Erode and Nilgiris. It has 104 affiliated institutions which consist of almost 29 colleges of education, 80 arts and science colleges, eight management institutions, one air force administration college and one college of physical education. The University opted for an evaluation by the National Assessment and Accreditation Council (NAAC) of the University Grants Commission. The apex council NAAC awarded "A" grade and subsequently re-accredited with "A" grade in 2009. The University has been ranked amongst top 50 Universities in India and placed at 32 and 33 in the surveys conducted by the popular English magazines India Today and The Week in 2013 respectively. In 2014, the University was ranked as 29th position by India Today magazine.

## REVIEW OF LITERATURE

Several manuscripts have been published in literature dealing with research performance of journals, countries, subjects and institutions. (Batcha, 2018)[1] discussed thoroughly about scientometric output of cardiovascular disease of SAARC countries and offers a powerful set of methods and measures for studying the structure and process of research communication. The paper examines the research trend, authorship, collaborative pattern and activity index of five SAARC Countries regarding the disease which amounts to about 24.8% of deaths in SAARC countries. The result of the paper reveals that India is a leader country among SAARC nations having major research output followed by Pakistan in cardiovascular disease research. The paper also deliberated that USA, England and Australia are the top collaboration countries which has done collaboration with SAARC nations.

(Sivakumar,2018)[2] studied and examined the analysis of research publications of the faculty of Bharathiar University, Coimbatore. Bibliographic records of 4645 items were retrieved from Scopus database between 1982 and 2015 and increasing publications trends were seen in Bharathiar University. The average output of the organization was 137 per year. Among the 4645 papers published in the span of 34 years, highest number of 748 papers was published in the year 2015. The total numbers of citation received by the papers were 50964 and the citing articles were 3559. The average citation per article was 14.3

(Batcha & Ahmad, 2017)[3] analysed comparative analysis of Indian Journal of Information Sources and Services (IJISS) and Pakistan Journal of Library and Information Science (PJLIS) during 2011-2017 and studied various aspects like year wise distribution of papers, authorship pattern & author productivity, degree of collaboration pattern of Co-Authorship , average length of papers , average keywords, etc and found 138(94.52%) of contributions from IJISS were made by Indian authors and similarly 94(77.05) of contributions from PJLIS were done by Pakistani authors. Papers by Indian and Pakistani Authors with Foreign Collaboration are minimal (1.37% of articles) and (4.10% of articles) respectively.

(Batcha, 2018)[4] analysed the various Scientometric components of the articles published by top six universities of Tamil Nadu from 2000-2017. The study identifies research trend, characteristics growth and collaboration pattern of published literature. The analysis of data reveals that the average growth rate increases at the rate of 9.76%. Further, the average citation per paper observed is 12.18%. High degree of international collaboration is notified and USA and South Korea are found to be the most preferred collaborative countries. The CAGR calculated for six universities are 9.76. The major research publications outputs are from the field of Chemistry, Crystallography and Pharmacy.

(Batcha, Jahina, & Ahmad, 2018)[5] has examined scientometric analysis of the DESIDOC Journal and analyzed the pattern of growth of the research output published in the journal, pattern of authorship, author productivity, and, subjects covered to the papers over the period (2013-2017). It found that 227 papers were published during the period of study (2001-2012). The maximum numbers of articles were collaborative in nature. The subject concentration of the journal noted was Scientometrics. The maximum numbers of articles (65 %) have ranged their thought contents between 6 and 10 pages.

(Maurya,2018)[6] examined and analyzed the scholarly communications in terms of scientometric assessment and presented the current scenario of Mizoram University's scholarly communications at the world level based on Web of Science database. Physics and Chemistry subjects found the highest productive area of scholarly communications, multiple authorship is prevalent with high degree of collaboration among authors, Department of Science and Technology (DST) as the topmost funding agency, South Korea found the strong collaborative country with Mizoram University in academic research, Tiwari D as the highest cited author, Thapa RK as the highest productive author, and increased growth in number of scholarly communications as well as citations.

(Batcha & Ahmad, 2017)[7] conducted scientometric analysis of 146 research articles published in Indian journal of Information Sources and Services (IJISS). The number of contributions, authorship pattern & author productivity, average citations, average length of articles, average keywords and collaborative papers was analyzed. Out of 146 contributions, only 39 were single authored and rest by multi authored with degree of collaboration 0.73 and week collaboration among the authors. The study concluded that the author productivity was 0.53 and was dominated by the Indian authors.

(Ahmad & Batcha,2019)[8] analyzed research productivity in Journal of Documentation (JDoc) for a period of 30 years between 1989 and 2018. Web of Science database a service from Clarivate Analytics has been used to download citation and source data. Bibexcel and Histcite application software have been used to present the datasets. Analysis part focuses on the parameters like citation impact at local and global level, influential authors and their total output, ranking of contributing institutions and countries. In addition to this scientographical mapping of data is presented through graphs using VOSviewer software mapping technique.

(Biljecki,2016)[9] examined a set of 12436 papers published in 20 GIScience journals in the gap of 2000-2014 and studied patterns and trends and its comprehensive scientometric study focuses on multiple aspects like output volume, citations, national output and efficiency, collaboration, altmetrics, authorship, and length of articles. The notable observation are that 5% countries contribute 76% global GIScience output, a paper published 15 years before received a median of 12 citations and the share of global collaborations in GIScience has more than tripled from the year 2000 onwards (31% papers has multiple authors from multiple countries in 2014 and it increased from 10% in 2014).

(Ahmad,et al, 2018)[10] explored scientometric analysis of the Webology Journal. The paper analyses the pattern of growth of the research output published in the journal, pattern of authorship, author productivity, and subjects covered to the papers over the period (2013-2017). It was found that 62 papers were published during the period of study (2013-2017). The maximum numbers of articles were collaborative in nature. The subject concentration of the journal noted was Social Networking/Web 2.0/Library 2.0 and Scientometrics or Bibliometrics. Iranian researchers contributed the maximum number of articles (37.10%). The study applied standard formula and statistical tools to bring out the factual results.

(Batcha, 2017)[11] analysed the research publication output in the field of robotic technology and shows that the robotic technology is a progressive field increasing the publication output from single digit to 513 year after year during the period from 1990 to 2016. The results shows that developing countries like USA, UK and Germany gives the most output on robotic technology related research. Yet major proportion of contribution (36.30%) is from USA. English language is the most preferred for the research amounting (87.70%) followed by German. The Prolific authors in the field of robotic technology are highly found from USA among them the contribution by Bloss R is appreciable and author from Japan, Dario P competes with more number of publications in the study.

## OBJECTIVES

- To examine the pattern of year wise growth of the research output of Bharathiar University.
- Identify the leading journals for publishing of scholarly communications from Bharathiar University.
- To identify the most preferred countries for collaboration for publishing their research results.
- Find out the top 30 prolific authors, collaborating institutions and collaborating countries.
- To determine publication density through mapping of top 20 authors, collaborating countries and institutions based on their number of research papers.

## METHODOLOGY

The data for the present study were downloaded from the Clarivate analytics-Web of Science database in July 2019. A total of 3440 research publications was downloaded from 2009-2018. The data downloaded were enhanced with different parameters like title, authors, years, collaborating countries, and research institutions. Furthermore, the downloaded data were analyzed by using Histcite, and VOSviewer software applications.

## RESULTS AND DISCUSSION

### Evaluate the Annual Output of Publications of Bharathiar University

The table I reveals that the numbers of research documents published from 2009 to 2018 are gradually increased. According to the publication output from the table I the year wise distribution of research documents, 2008 has the highest number of research documents 603 (17.73%) with 94 (2.17%) of total local citation score and 1371 (3.62%) of total global citation score values and being prominent among the 10 years output and it stood in first rank position. The year 2017 has 574 (16.69%) research documents and it stood in second position with 344 (7.92%) of total local citation score and 3184 (8.40%) of total global citation score were scaled. It is followed by the year 2016 with 449 (13.05 %) of records and it stood in third rank position along with 494 (11.38%) of total local citation score and 4107 (10.83%) of total global citation score measured. The year 2015 has 381 (11.08%) research documents and it stood in fourth position with 779 (17.95%) of total local citation score and 5214 (13.75%) of total global citation score were scaled. It is noticed that the increase in publications may not create impact on citation score yet the quality matters on total local citation scores and on total global citation scores. It clearly indicates on the fact that the increased publication rate is not bringing the increased citation rate.

**Table1.** *Annual Distribution of Publications and Citations*

| S. No. | Year | Records | % | Rank | TLCS | % | TGCS | % |
|---|---|---|---|---|---|---|---|---|
| 1 | 2009 | 135 | 3.92 | 10 | 165 | 3.80 | 1977 | 5.22 |
| 2 | 2010 | 194 | 5.64 | 9 | 226 | 5.21 | 3290 | 8.68 |
| 3 | 2011 | 212 | 6.16 | 8 | 526 | 12.12 | 3807 | 10.04 |
| 4 | 2012 | 260 | 7.56 | 7 | 738 | 17.00 | 5077 | 13.39 |
| 5 | 2013 | 281 | 8.17 | 6 | 472 | 10.87 | 4972 | 13.12 |
| 6 | 2014 | 351 | 10.20 | 5 | 503 | 11.59 | 4909 | 12.95 |
| 7 | 2015 | 381 | 11.08 | 4 | 779 | 17.95 | 5214 | 13.75 |
| 8 | 2016 | 449 | 13.05 | 3 | 494 | 11.38 | 4107 | 10.83 |
| 9 | 2017 | 574 | 16.69 | 2 | 344 | 7.92 | 3184 | 8.40 |
| 10 | 2018 | 603 | 17.53 | 1 | 94 | 2.17 | 1371 | 3.62 |
| **Total** | | **3440** | 100.00 | | 4341 | 100.00 | 37908 | 100.00 |

*#TLCS = Total Local Citation Score, #TGCS = Total Global Citation Score*

## Analysis of the Publication Output of Top 20 Authors of Bharathiar University

Table II and figure 1 displays the ranking of authors of research articles. In the rank analysis the authors who have published 50 articles or more are considered into account to avoid a long list. It was observed that there is total of 5464 authors for 3440 records and it shows the top 20 most productive authors during 2009-2018. Murugan K published 173 (5.03%) articles with 3137 TGCS articles, followed by Rakkiyappan R 133 (3.87%) with 3356 TGCS articles, Mangalaraj D 126 (3.66%) with 2599 TGCS articles, Huang CY 113 (3.28%) with 589 TGCS articles, Balachandran K 109 (3.17%) with 1358 TGCS article, Benelli G 100 (2.91%) with 1988 TGCS articles, other authors have contributed less than 3% during the period of study. The data set clearly depicts that no matter how many publications that an author brings out yet the quality publications alone shows impact in the form of total local citations score and total global citations score. It could be identified that the authors' wise analysis the following authors Murugan K, Rakkiyappan R, Mangalaraj D, Huang CY, Balachandran K, and Benelli G, were identified the most productive authors based on the number of research papers published. The data set puts forth that the authors Rakkiyappan R with 3356 citations, Murugan K with 3137 citations, Selvan RK with 2839 citations and Mangalaraj D with 2599 citations.

**Table2.** *Publication output of Top 20 Authors and Citation Score*

| S. No. | Authors | Records | % | TLCS | TGCS |
|---|---|---|---|---|---|
| 1 | Murugan K | 173 | 5.03 | 1052 | 3137 |
| 2 | Rakkiyappan R | 133 | 3.87 | 137 | 3356 |
| 3 | Mangalaraj D | 126 | 3.66 | 183 | 2599 |
| 4 | Huang CY | 113 | 3.28 | 84 | 589 |
| 5 | Balachandran K | 109 | 3.17 | 199 | 1358 |
| 6 | Benelli G | 100 | 2.91 | 664 | 1988 |
| 7 | Ponpandian N | 91 | 2.65 | 151 | 1889 |
| 8 | Selvan RK | 91 | 2.65 | 234 | 2839 |
| 9 | Prasad KJR | 89 | 2.59 | 154 | 454 |
| 10 | Kolandaivel P | 84 | 2.44 | 181 | 682 |
| 11 | Padma VV | 82 | 2.38 | 248 | 1443 |
| 12 | Senthilkumar K | 68 | 1.98 | 113 | 582 |
| 13 | Natarajan K | 64 | 1.86 | 408 | 2091 |
| 14 | Kuo WW | 63 | 1.83 | 55 | 365 |
| 15 | Viswanadha VP | 59 | 1.72 | 41 | 321 |
| 16 | Panneerselvam C | 57 | 1.66 | 709 | 1795 |
| 17 | Nicoletti M | 55 | 1.60 | 646 | 1685 |
| 18 | Prabhakaran R | 52 | 1.51 | 284 | 1028 |
| 19 | Higuchi A | 51 | 1.48 | 258 | 840 |
| 20 | Nataraj D | 50 | 1.45 | 64 | 1003 |
| | **Total** | 1710 | 49.71 | 5865 | 30044 |

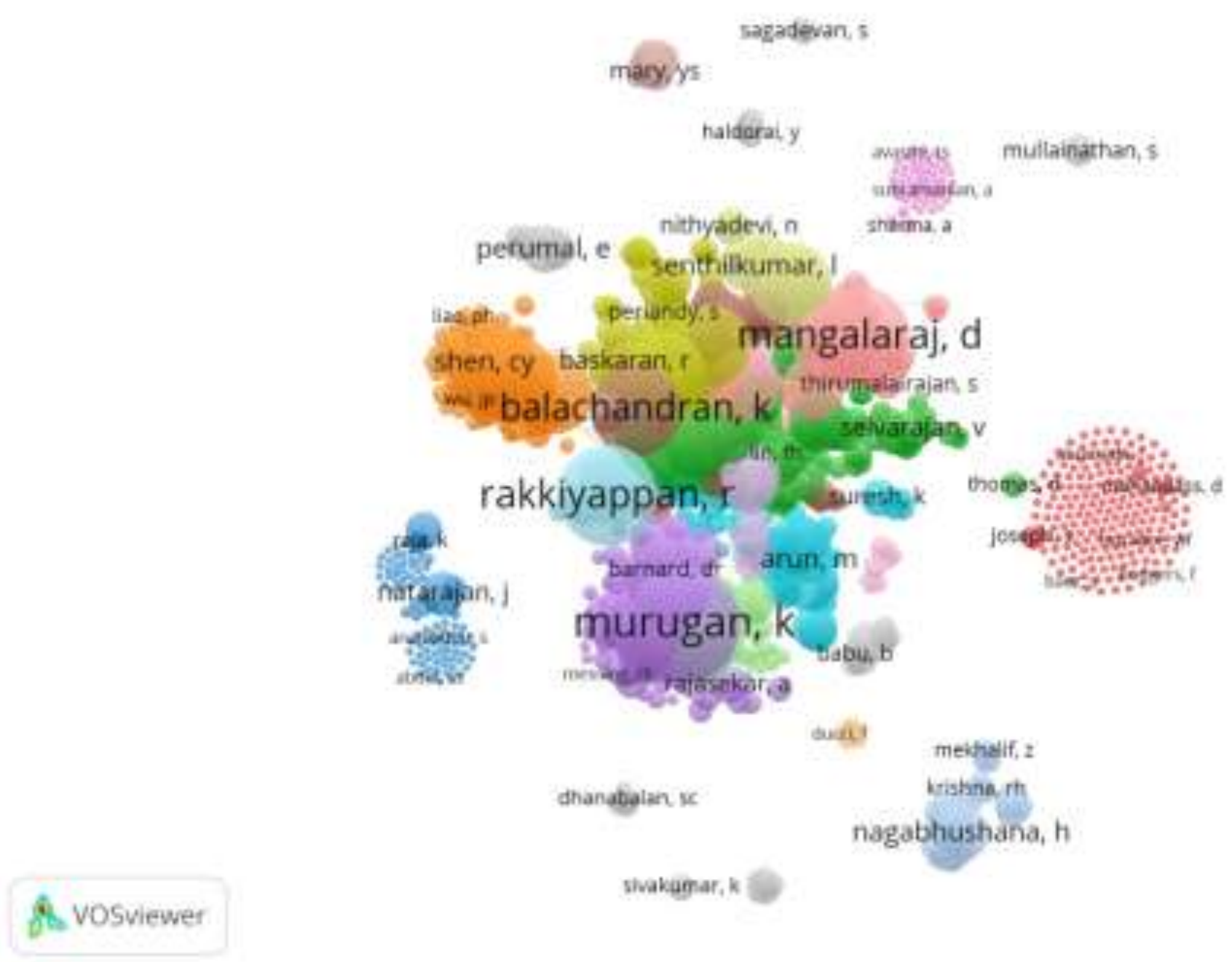

**Figure1.** *Showing Highly Prolific authors*

## Journal Wise Distribution of Documents

The rankings of top 20 journal wise distribution of documents are depicted in the following Table III.

**Table3.** *Top 20 Journal Wise Distribution of Documents*

| S. No. | Journals | Documents | % | TLCS | TGCS |
|---|---|---|---|---|---|
| 1 | RSC Advances | 88 | 2.56 | 187 | 1582 |
| 2 | Spectrochimica Acta Part A-Molecular And Biomolecular Spectroscopy | 75 | 2.18 | 64 | 1080 |
| 3 | Journal Of Molecular Structure | 57 | 1.66 | 48 | 357 |
| 4 | Parasitology Research | 52 | 1.51 | 669 | 1734 |
| 5 | Journal Of Materials Science-Materials In Electronics | 43 | 1.25 | 13 | 211 |
| 6 | Journal Of Alloys And Compounds | 29 | 0.84 | 58 | 585 |
| 7 | Optik | 29 | 0.84 | 15 | 82 |
| 8 | Synthetic Communications | 29 | 0.84 | 39 | 105 |
| 9 | Ionics | 28 | 0.81 | 42 | 270 |
| 10 | Inorganica Chimica Acta | 27 | 0.78 | 83 | 442 |
| 11 | Materials Research Bulletin | 25 | 0.73 | 35 | 377 |
| 12 | Neurocomputing | 24 | 0.70 | 18 | 669 |
| 13 | Journal Of Cluster Science | 23 | 0.67 | 26 | 134 |
| 14 | New Journal Of Chemistry | 23 | 0.67 | 31 | 152 |
| 15 | Biomedicine & Pharmacotherapy | 22 | 0.64 | 13 | 110 |
| 16 | Environmental Science And Pollution Research | 22 | 0.64 | 76 | 260 |
| 17 | International Journal Of Biological Macromolecules | 22 | 0.64 | 14 | 277 |
| 18 | Applied Surface Science | 21 | 0.61 | 39 | 481 |
| 19 | Environmental Toxicology | 21 | 0.61 | 15 | 97 |
| 20 | Acta Crystallographica Section E-Structure Reports Online | 20 | 0.58 | 11 | 28 |

Table III reveals that RSC Advances published 88 (2.56%) papers, Spectrochimica Acta Part A-Molecular and Biomolecular Spectroscopy published 75 papers, (2.18%), Journal Of Molecular Structure published 57 (1.66%) papers and Parasitology Research published 52 (1.51%) papers.

## Ranking of Collaborative Institutions

The productivity of the author publications based on the collaborative institutions is depicted in the following Table IV.

**Table4.** *Ranking of Collaborative Institutions*

| S. No. | Institution | Records | % | TLCS | TGCS | ACPP |
|---|---|---|---|---|---|---|
| 1 | Asia University | 107 | 3.11 | 73 | 563 | 5.26 |
| 2 | China Medical University | 107 | 3.11 | 73 | 563 | 5.26 |
| 3 | University Pisa | 102 | 2.97 | 664 | 1996 | 19.57 |
| 4 | King Saud University | 97 | 2.82 | 327 | 1504 | 15.51 |
| 5 | Periyar University | 81 | 2.35 | 115 | 916 | 11.31 |
| 6 | Annamalai University | 72 | 2.09 | 47 | 768 | 10.67 |
| 7 | Bharathidasan University | 71 | 2.06 | 154 | 1045 | 14.72 |
| 8 | Thiruvalluvar University | 68 | 1.98 | 104 | 565 | 8.31 |
| 9 | Govt Arts College | 58 | 1.69 | 20 | 267 | 4.60 |
| 10 | Anna University | 51 | 1.48 | 16 | 386 | 7.57 |
| 11 | China Medical University Hospital | 49 | 1.42 | 43 | 257 | 5.24 |
| 12 | Changhua Christian Hospital | 45 | 1.31 | 32 | 228 | 5.07 |
| 13 | Karpagam University | 44 | 1.28 | 19 | 360 | 8.18 |
| 14 | Natl Cent University | 42 | 1.22 | 97 | 431 | 10.26 |
| 15 | Sungkyunkwan University | 42 | 1.22 | 24 | 245 | 5.83 |
| 16 | University Putra Malaysia | 39 | 1.13 | 120 | 500 | 12.82 |
| 17 | Karunya University | 38 | 1.10 | 19 | 308 | 8.11 |
| 18 | King Abdulaziz University | 38 | 1.10 | 45 | 1064 | 28.00 |
| 19 | Sri Ramakrishna Mission Vidyalaya College Arts & Science | 37 | 1.08 | 15 | 242 | 6.54 |
| 20 | VIT University | 37 | 1.08 | 27 | 320 | 8.65 |

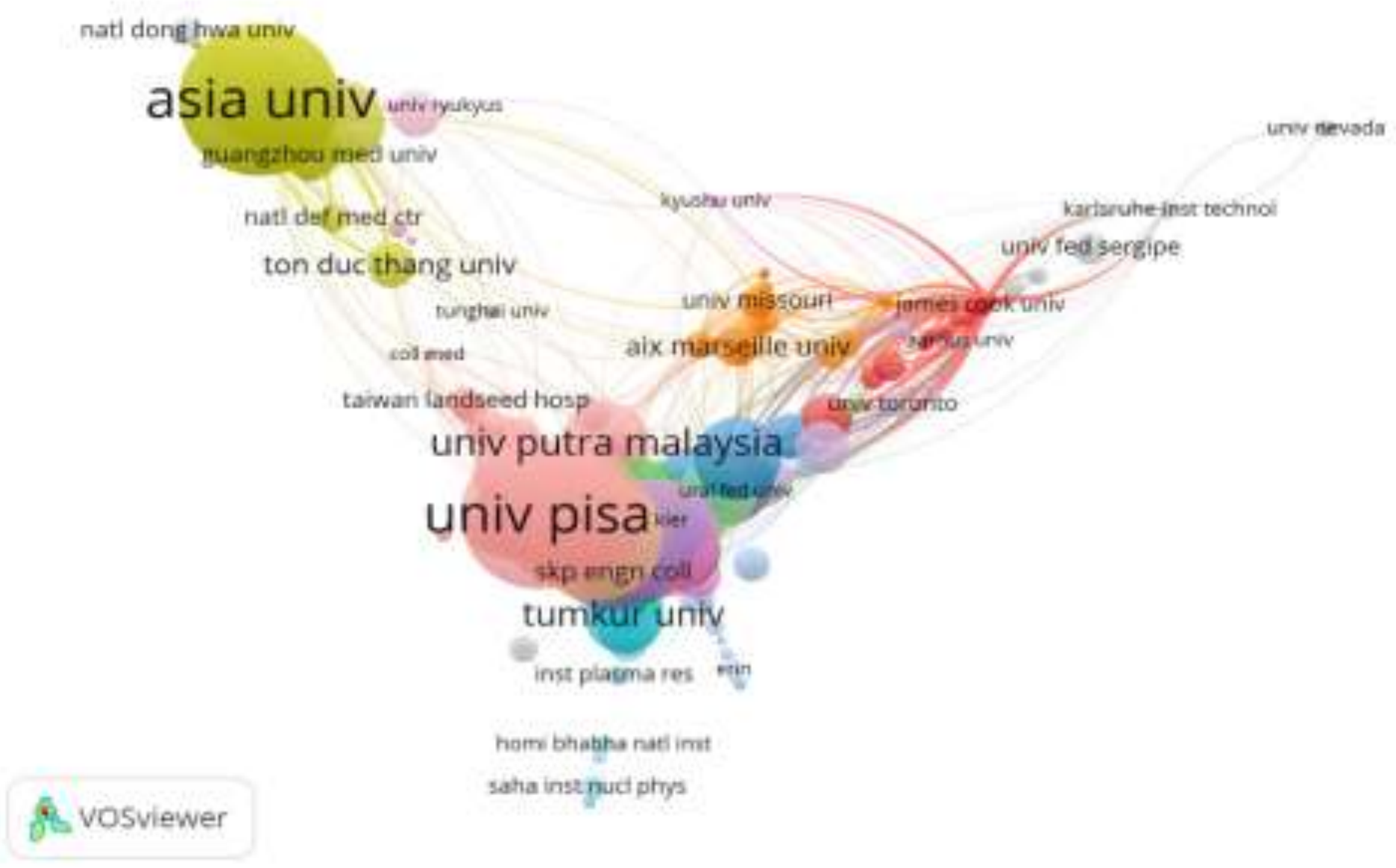

**Figure2.** *Collaboration of Institutions and their clusters*

It is found that in total 1851 institutions, including 3436 subdivisions published 3440 research papers during 2009 - 2018 and these institutions have collaborated with faculty of Bharathiar University for research and publications. The topmost twenty prolific institutions involved in this research have published 37 and more research articles. The mean average is 1.86 research articles per institution. Out of 1851 institutions, top 20 institutions published 1225 (35.61%) research papers and the rest of the institution published 2215 (64.39%) research papers respectively. Based on the number of published research records the institutions are ranked. Table IV summarizes articles, the global citation score, local citation score and average citation per paper of the publications of these institutions.

## Ranking of Department Wise Distribution

The ranking based on the Department wise distribution of the Bharathiar University is depicted in the following Table V.

**Table5.** *Ranking of Department Wise Distribution.*

| S. No. | Department of University | Records | % | TLCS | TGCS | ACPP |
|---|---|---|---|---|---|---|
| 1 | Department of Physics | 572 | 1.66 | 786 | 7305 | 12.77 |
| 2 | Centre for Research & Development | 382 | 1.11 | 322 | 3239 | 8.48 |
| 3 | Department of Chemistry | 370 | 1.08 | 913 | 5049 | 13.65 |
| 4 | Department of Mathematics | 364 | 1.06 | 408 | 5608 | 15.41 |
| 5 | Department of Biotechnology | 236 | 0.69 | 322 | 2709 | 11.48 |
| 6 | Department of NanoScience & Technology | 233 | 0.68 | 245 | 3785 | 16.24 |
| 7 | Research & Development Centre | 204 | 0.59 | 67 | 588 | 2.88 |
| 8 | School of Life Science | 189 | 0.55 | 778 | 2911 | 15.40 |
| 9 | Department of Zoology | 153 | 0.44 | 343 | 1909 | 12.48 |
| 10 | Department of Botany | 130 | 0.38 | 62 | 840 | 6.46 |
| 11 | Department of Environmental Science | 70 | 0.20 | 25 | 844 | 12.06 |
| 12 | Department of Microbial Biotechnology | 70 | 0.20 | 26 | 627 | 8.96 |
| 13 | Department of Bioinformatics | 62 | 0.18 | 67 | 521 | 8.40 |
| 14 | DRDO BU Centre for Life Science | 50 | 0.15 | 112 | 682 | 13.64 |
| 15 | Department of Applied Mathematics | 35 | 0.10 | 20 | 208 | 5.94 |
| 16 | R&D Centre | 34 | 0.10 | 12 | 113 | 3.32 |
| 17 | School of Biotechnology & Genetic Engineering | 29 | 0.08 | 39 | 391 | 13.48 |
| 18 | Department of Medical Physics | 20 | 0.06 | 3 | 59 | 2.95 |
| 19 | Department of Computer Science | 19 | 0.06 | 1 | 38 | 2.00 |
| 20 | School of Physical Science | 19 | 0.06 | 10 | 179 | 9.42 |

Table V shows the department wise distribution. There are 572 articles (1.66%) which were published by the authors of Bharathiar University from department of Physics followed by Centre for Research & Development with 382 articles (1.11%), Department of Chemistry with 370 articles (1.08%), Department of Mathematics with 364 articles (1.06%) and the rest of the departments published less than one percent of research articles from 2009 to 2018.

## Country Wise Collaborations

Table VI and figure 3 displays that faculty of Bharathiar University has collaborated with South Korea 289 (8.40%) publications followed by Taiwan with 231 (6.72%) publications and USA with 222 (6.45%) publications. It also shows that Bharathiar University has collaborated and contributed good number of papers with Peoples Republic of China, Saudi Arabia, Italy and Japan.

**Table6.** *Country Wise Collaborations*

| S. No. | Countries | Records | % | TLCS | TGCS |
|---|---|---|---|---|---|
| 1 | South Korea | 289 | 8.40116 | 262 | 4477 |
| 2 | Taiwan | 231 | 6.71512 | 545 | 2759 |
| 3 | USA | 222 | 6.45349 | 761 | 4102 |
| 4 | Peoples Republic China | 183 | 5.31977 | 305 | 3562 |
| 5 | Saudi Arabia | 164 | 4.76744 | 420 | 2836 |
| 6 | Italy | 129 | 3.75 | 711 | 2341 |
| 7 | Japan | 80 | 2.32558 | 261 | 1403 |
| 8 | Malaysia | 63 | 1.8314 | 127 | 646 |
| 9 | Canada | 58 | 1.68605 | 47 | 915 |
| 10 | Australia | 50 | 1.45349 | 43 | 978 |
| 11 | Germany | 47 | 1.36628 | 92 | 598 |
| 12 | France | 44 | 1.27907 | 214 | 771 |
| 13 | UK | 44 | 1.27907 | 59 | 343 |
| 14 | Singapore | 40 | 1.16279 | 31 | 980 |
| 15 | Turkey | 40 | 1.16279 | 29 | 390 |
| 16 | Spain | 35 | 1.01744 | 65 | 560 |
| 17 | United Arab Emirates | 31 | 0.90116 | 24 | 463 |
| 18 | Brazil | 30 | 0.87209 | 22 | 493 |
| 19 | Belgium | 29 | 0.84302 | 5 | 231 |
| 20 | South Africa | 25 | 0.72674 | 18 | 143 |

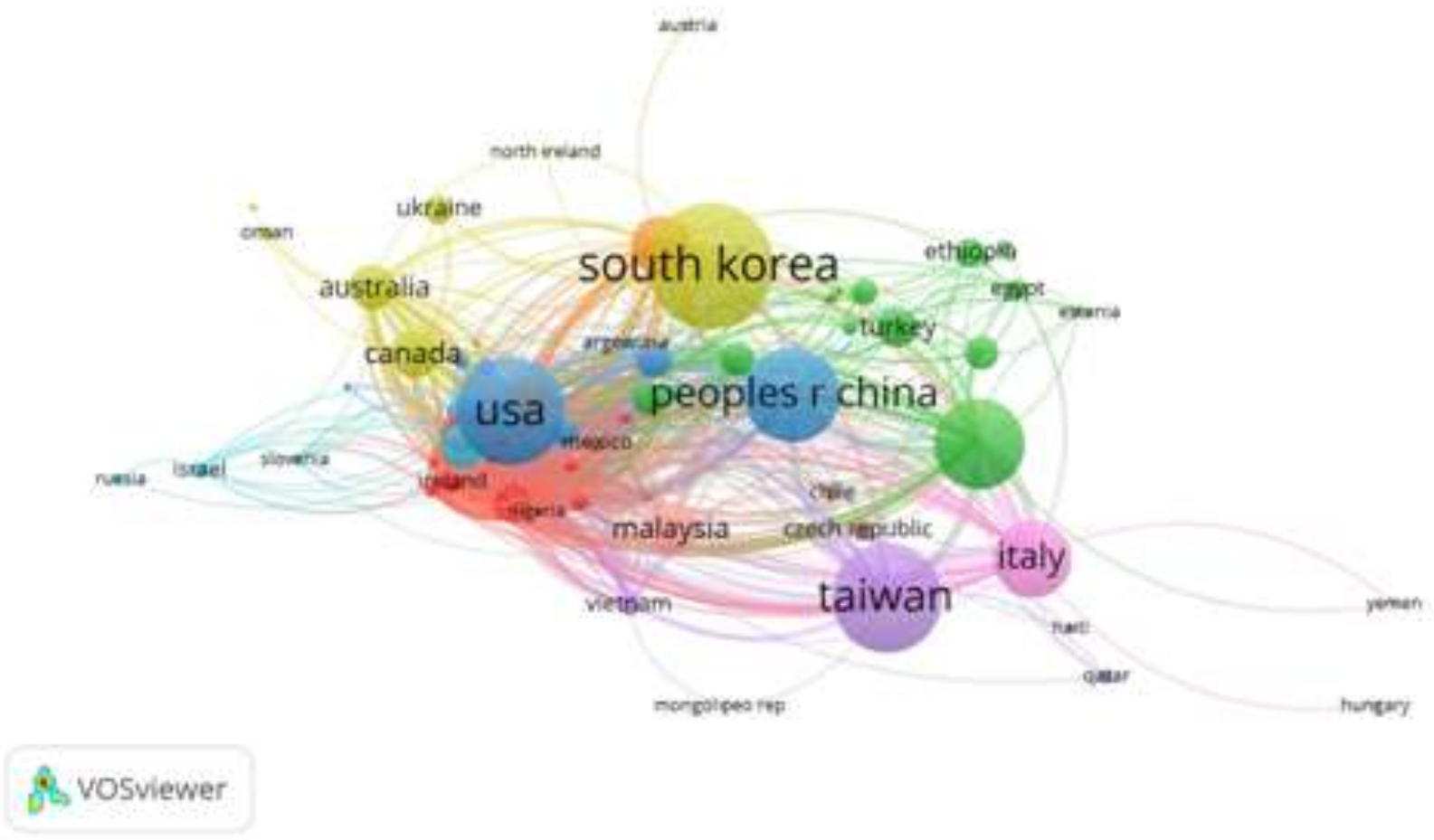

**Figure3.** *Showing Ranking of Country wise Collaboration*

## CONCLUSION

Research productivity in the Bharathiar University among the faculty is significantly high. Though the study started in recent decade but there is really an optimistic growth in the research productivity. Production is the real asset for any institution but as compared to other organization/institution, still, Bharathiar University needs to improve the research performance to a higher level to meet or be at par with leading Universities.